\documentclass[lettersize,journal]{IEEEtran}
\usepackage[utf8]{inputenc}
\usepackage{amsfonts}
\usepackage{amssymb}
\usepackage{amsmath}
\usepackage{microtype}
\usepackage{graphicx}
\usepackage{booktabs}
\usepackage{hyperref}
\usepackage{multirow}
\usepackage{footnote}
\usepackage{tablefootnote}
\usepackage{balance}
\usepackage{siunitx} 
\usepackage{xcolor} 
\usepackage{gensymb} 
\usepackage{commath} 
\usepackage{bm} 
\usepackage[noadjust]{cite} 
\usepackage{environ} 
\usepackage{breqn} 
\usepackage{threeparttable} 
\usepackage{algorithm} 
\usepackage{algpseudocode}

\makesavenoteenv{table}
\makesavenoteenv{table*}
\makesavenoteenv{tabular}

\newcommand{\R}[1] {\textnormal{#1}}

\NewEnviron{NORMAL}{%
    \scalebox{1}{$\BODY$} 
} 

\begin{document}

\title{Static Background Removal in Vehicular Radar: Filtering in Azimuth-Elevation-Doppler Domain}

\author{Xiangyu Gao, Sumit Roy,~\IEEEmembership{Fellow,~IEEE}, Lyutianyang Zhang
\thanks{X. Gao, S. Roy, and L. Zhang are with the Department of Electrical and Computer Engineering, University of Washington, Seattle, WA, 98195, USA. E-mail: xygao@uw.edu, sroy@uw.edu, lyutiz@uw.edu.}
}

\maketitle

\begin{abstract}
Anti-collision assistance, integral to the current drive towards increased vehicular autonomy, relies heavily on precise detection and localization of moving targets in the vehicle's vicinity. A crucial step towards achieving this is the removal of static objects from the scene, thereby enhancing the detection and localization of dynamic targets—a pivotal aspect in augmenting overall system performance. In this paper, we propose a static background removal algorithm tailored for automotive scenarios, designed for common frequency-modulated continuous wave (FMCW) radars. This algorithm effectively eliminates reflections corresponding to static backgrounds from radar images through a two-step process: 4-dimensional (4D) radar imaging and filtering in the azimuth-elevation-Doppler domain. Our proposed approach is underpinned by a model customized for FMCW radar signals, incorporating a time-division multiplexing-based multiple-input multiple-output scheme on the non-uniform radar array. Furthermore, our filtering process requires knowledge of the 3-dimensional (3D) radar ego-motion velocity, typically obtained from an external sensor. To address scenarios where such sensors are unavailable, we introduce a self-contained 3D ego-motion estimation approach. Finally, we evaluate the performance of our algorithm using both simulated and real-world data, analyzing its sensitivity and time complexity in comparison to established baselines.
\end{abstract}


\section{Introduction}
Autonomous driving systems that rely on multi-sensor fusion and scene perception are key to achieving future L4 and L5 vehicular automation \cite{ramp, 10008067, mimosar, sihao2023deformable, 9319548, xiangyu2022deformable}. However, image understanding in complex environments, such as city roads with dense and varied traffic, remains a significant challenge. In these scenarios, which involve both moving and static objects, removing static elements is a common method to enhance moving target indication (MTI) \cite{melvin2004stap}.

\subsection{Related Works}
MTI, originally developed for airborne radar systems \cite{xu2014space}, is used to detect and track moving targets by filtering out clutter—unwanted echoes from stationary objects. To improve MTI in air-to-air and air-to-ground scenarios, space-time adaptive processing (STAP) is a well-established technique \cite{xu2014space}. STAP uses 2D joint adaptive filtering in both spatial and temporal domains to maximize the signal-to-interference ratio (SIR) \cite{melvin2004stap}. However, optimal STAP filters require prior knowledge of clutter statistics in the test area and a sufficient number of independent and identically distributed (IID) training samples, which is often challenging in real-time vehicular radar applications due to varying objects and backgrounds \cite{gu2022end}.

Beyond STAP-based approaches, several methods focus on clutter suppression in automotive radar systems. Yoon et al. \cite{yoon2019high} propose analyzing beat frequency distributions in FMCW radar signals to recognize and mitigate clutter, especially from structures like guardrails and tunnels. Similarly, Lee et al. \cite{lee2017harmonic} explore radar signal periodicity to address harmonic clutter caused by tunnels and soundproof walls. Matsunami et al. \cite{matsunami2010clutter} introduce a clutter suppression method based on pulse integration and target occurrence probability for detecting multiple vehicles. Yu et al. \cite{yu2012mimo} propose a MIMO beamforming technique to mitigate multipath clutter from large specular reflectors such as highway guardrails or buildings.

In ground-penetrating and through-wall radars, background removal is used to filter out clutter from ground or walls, revealing hidden targets \cite{6689119}. Techniques such as coherent background subtraction, mean subtraction \cite{5505008}, frame differencing \cite{9337322}, and singular-value decomposition (SVD) \cite{tivive2011svd} have been applied. However, coherent background subtraction, which requires knowledge of the wall characteristics, is less feasible in automotive contexts. Frame differencing and mean subtraction leverage the time and angle invariance of clutter to suppress it.

Deep learning (DL) \cite{ke2020enhancing, ke2024interpretable, zheng2023coordinated, zheng2024enhancing, ke2024real} has also contributed significantly to clutter suppression, with DL frameworks applied to automotive radar image processing \cite{gao2019experiments, 9765320} and airborne radars. Gu et al. \cite{gu2022end} developed a DL framework to address clutter suppression in non-homogeneous environments, dealing with issues like insufficient training data and low detection probabilities. Additionally, several open-source datasets \cite{xm40-jx59-22, begn-ye78-22}, collected using \SI{77}{GHz} FMCW radar, are available for testing radar algorithms in real-world driving or indoor scenarios.

\subsection{Contributions}
Recent advancements in \SI{77}{GHz} FMCW radars have demonstrated highly accurate object detection and localization, regardless of environmental conditions \cite{gao2019experiments, 9695280, gao2021perception, ramp}. These radars are particularly effective at fine-resolution Doppler velocity measurements \cite{gao2019experiments}, making them ideal for background removal in automotive scenarios. The Doppler velocities of static objects are determined by their azimuth and elevation angles, as well as the radar's instantaneous velocity and direction \cite{mimosar}. By identifying the specific Doppler velocity profile of background objects, we propose to filter out clutter in the azimuth-elevation-Doppler domain using notch filtering. This approach is more efficient than traditional temporal and spatial filtering methods, as it requires no training data, is applicable to both moving and stationary radar, and avoids complex STAP-like covariance matrix computations.

Our contributions are summarized as follows:
\begin{itemize}
\item We developed a model tailored for FMCW radar signals, optimized for point target detection in automotive scenarios. To enhance resolution, we integrated a non-uniform radar array with a time-division multiplexing-based MIMO scheme.
\item Our static background removal algorithm operates in two stages. First, we combine conventional range and Doppler processing with subarray-based azimuth and elevation processing to reduce sidelobes from the non-uniform array. Second, we apply notch filtering to remove Doppler frequencies associated with background clutter.
\item We introduce a self-contained 3D radar ego-motion estimation method for background removal, utilizing constant-false-alarm-rate (CFAR) target detection to generate radar point clouds. This method also addresses Doppler ambiguity issues common in large MIMO systems.
\item We evaluated the performance of our algorithms using simulated data and real-world experiments with data collected from Texas Instrument’s cascaded-chip radar board in practical driving scenarios.
\end{itemize}

The remainder of this paper is structured as follows: Section II introduces the signal model for vehicular FMCW MIMO radar. Sections III and IV describe the proposed algorithm in detail. Section V presents the simulation results, while Section VI discusses experimental results and analysis. Finally, Section VII concludes the paper.

\section{Signal Model}
\subsection{FMCW Signal for A Vehicular Radar}
In this section, we model a frame of the FMCW radar return signal for a general point target in autonomous scenarios. We assume that each radar transmitter (TX) transmits a sequence of $N_{\R{c}}$ chirps with duration $T_{\R{c}}$ in a frame. With carrier frequency $f_{\R{c}}$ and chirp slope $S_{\R{w}}$, transmitter power $A_{\R{T}}$, and initial phase $\phi_0$, the transmit signal of the FMCW radar during time $t$ within a frame is given by \cite{mimosar}:
\begin{equation}
    s_{\R{T}} (t)=A_{\R{T}}  \cos \left(2\pi \left(f_{\R{c}} t+\frac{1}{2}S_{\R{w}} t^2 \right) + \phi_0 \right).
\end{equation}

For target at distance $r$ from radar with reflection coefficient $\beta$, the received reflected signal $s_{\R{R}}(t)$ incurs round-trip delay $\tau=2r/c_0$,  i.e., $s_{\R{R}}(t)= A_{\R{T}}A_{\R{R}} \beta s_{\R{T}} (t-\tau)$, where $c_0$ is the speed of light and $A_{\R{R}}$ is the receiver power \cite{gao2019experiments}. The received signal is then mixed with the transmit signal at the receiver to produce the difference intermediate frequency (IF) signal $s_{\R{IF}}(t)$:
\begin{dmath}
    s_{\R{IF}}(t) = \frac{A_{\R{T}} A_{\R{R}}\beta}{2} \Bigg\{ \cos \left[2 \pi\left(S_{\R{w}} \tau t+f_{\R{c}} \tau-\frac{1}{2} S_{\R{w}} \tau^{2}\right)\right] \Bigg\}.
 \label{eq:sif}
\end{dmath}

\begin{figure}
\centering
\includegraphics[width=0.45\textwidth]{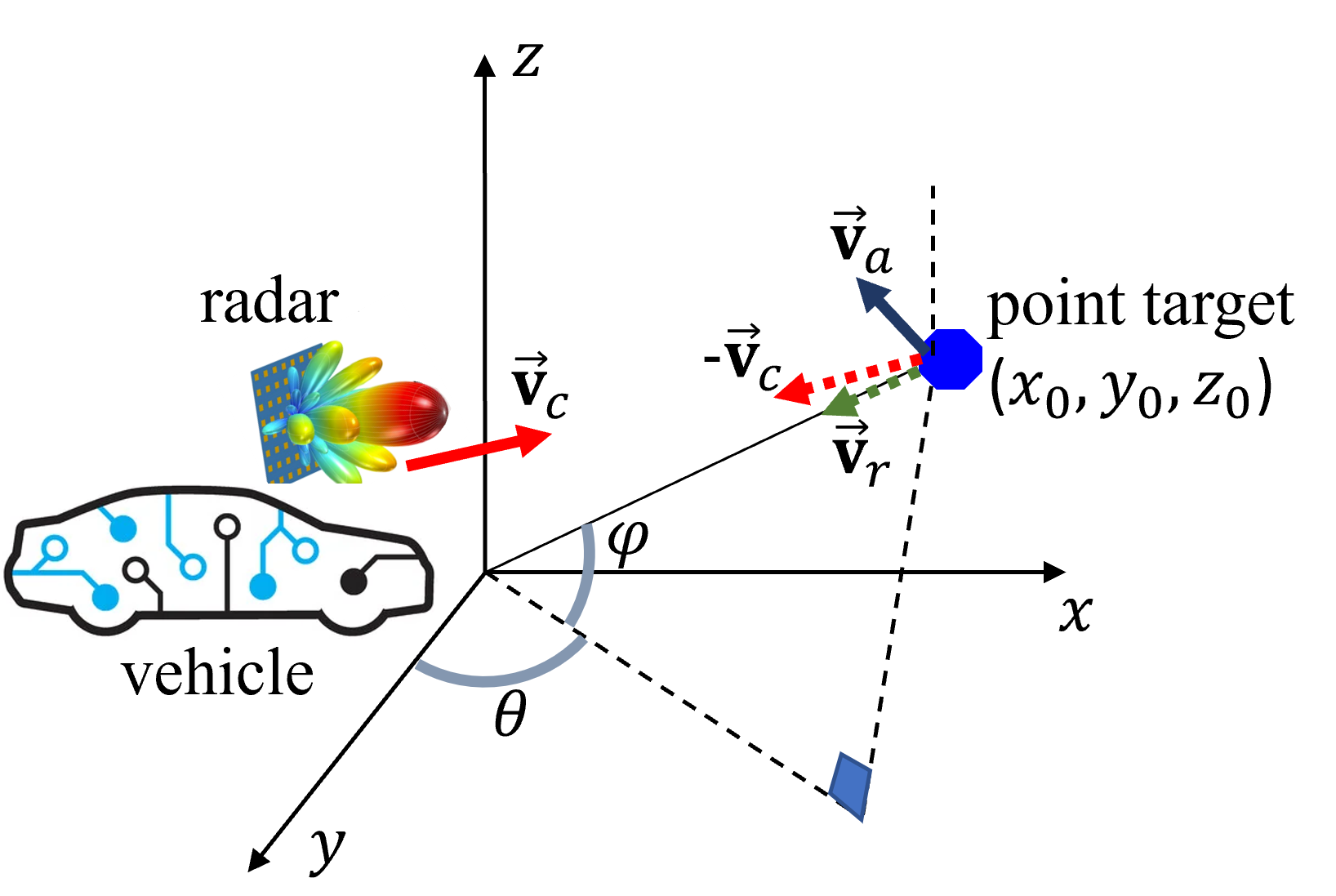}
\caption{Model scenario - a moving radar and a moving point target in global Cartesian coordinates for a single frame at $t=0$, with radar at origin and target at $(x_0, y_0, z_0)$. Radar (target) moves with velocity $\vec{\mathbf{v}}_c$ ($\vec{\mathbf{v}}_a$), respectively. The target exhibits a relative Doppler velocity $\vec{\mathbf{v}}_r$ with respect to the radar over this frame.}
\label{fig:sentar}
\end{figure}

We assume the vehicle-mounted radar moves in the global coordinate system shown in Fig.~\ref{fig:sentar}, from the origin at $t=0$ with ego velocity $\vec{\mathbf{v}}_c(t)=\left(v_x(t), v_y(t), v_z(t)\right)$ at time $t$. For frame-by-frame modeling, we assume that the velocities of the radar and any targets may be assumed constant over the short frame duration (typically milliseconds). Therefore, we simplify the time-dependant variables $\vec{\mathbf{v}}_c(t)$ by $\vec{\mathbf{v}}_c$ in the following analysis. The point target is located at a range $r$, azimuth angle $\theta$, and elevation angle $\varphi$, with corresponding Cartesian-coordinates location $(x_0, y_0, z_0)=(r\cos{\varphi}\sin\theta, r\cos \varphi\cos \theta, r\sin \varphi)$, moving with velocity $\vec{\mathbf{v}}_a=(v_{a, x}, v_{a, y}, v_{a, z})$. The target would exhibit a relative Doppler velocity (the velocity along the radial direction) $\vec{\mathbf{v}}_r$ with respect to the radar; the amplitude of $\vec{\mathbf{v}}_r$ (denoted by $v_r$) is obtained by projecting the inverse platform velocity $-\vec{\mathbf{v}}_c$ and target velocity $\vec{\mathbf{v}}_a$ onto the radial direction:
\begin{equation}
\label{eq:vr_mov}
\begin{aligned}
    v_r = & \left((v_{a, y}-v_{y})\cos{\theta}+(v_{a, x}-v_{x})\sin{\theta}\right)\cos{\varphi} \\ & + (v_{a, z}-v_{z})\sin{\varphi}.
\end{aligned}
\end{equation}

If $\vec{\mathbf{v}}_a=\mathbf{0}$ (i.e., the target is stationary), the relative Doppler velocity amplitude is given by:
\begin{equation}
    v_r = -(v_{y}\cos{\theta}+v_{x}\sin{\theta})\cos{\varphi} -v_{z}\sin{\varphi}.
\label{eq:vr}
\end{equation}

For measuring the Doppler velocity, FMCW radar sends a sequence of chirps within a frame, as the round-trip delay at each chirp varies slightly due to relative motion. By decomposing the $t$ into fast time $t_f$ (i.e., time within a chirp) and slow time $n$ (the chirp index), i.e., $t=nT_c+t_f$, we can represent the round-trip delay for the target at $n$-th chirp $\tau_n$ using its Doppler velocity $v_r$ and the round-trip delay for the first chirp $\tau_0$:
\begin{equation}
    \tau_n = \tau_{0} - \frac{2 v_{\R{r}} n T_{\R{c}}}{c_0} =  \tau_{0} + n \Delta \tau,
\label{eq:tau}
\end{equation}
\noindent where $\Delta \tau=-\frac{2 v_{\R{r}} T_{\R{c}}}{c_0}$, and $\tau_0=\frac{2\sqrt{x_0^2+y_0^2+z_0^2}}{c_0}$.

Substituting Eq.~\eqref{eq:tau} into Eq.~\eqref{eq:sif} and then sampling in fast time $t_f$ with frequency $f_{\R{s}}$, we get the post-analog-to-digital conversion (ADC) sampled IF signal for any chirp $n$ and fast-time sample $m$:
\begin{equation}
    s_{\R{IF}}(m,n)=\frac{A_{\R{T}} A_{\R{R}}\beta}{2} \exp \left(j 2 \pi\left(S_{\R{w}} \tau_n \frac{m}{f_{\R{s}}} + f_{\R{c}} \tau_n - \frac{1}{2} S_{\R{w}} \tau_n^{2}\right)\right).
    \label{eq:ifs_dop}
\end{equation}

\subsection{Non-uniform Radar Array and MIMO Signal}
To enhance resolution, state-of-the-art radars typically employ multiple transmitters and receivers to create a larger aperture. In the automotive radar domain, a common approach involves uniform spacing of transmit-and-receive arrays, referred to as a ``uniform array". However, non-uniform arrays, such as the minimum redundancy array (MRA) \cite{mra}, can achieve larger apertures with the same number of antennas, albeit at the cost of some empty array elements \cite{gao2021perception}. Therefore, we propose the utilization of a non-uniform array in our radar system and proceed to model its MIMO signal in the subsequent sections.

\begin{figure}
\centering
\includegraphics[width=0.48\textwidth]{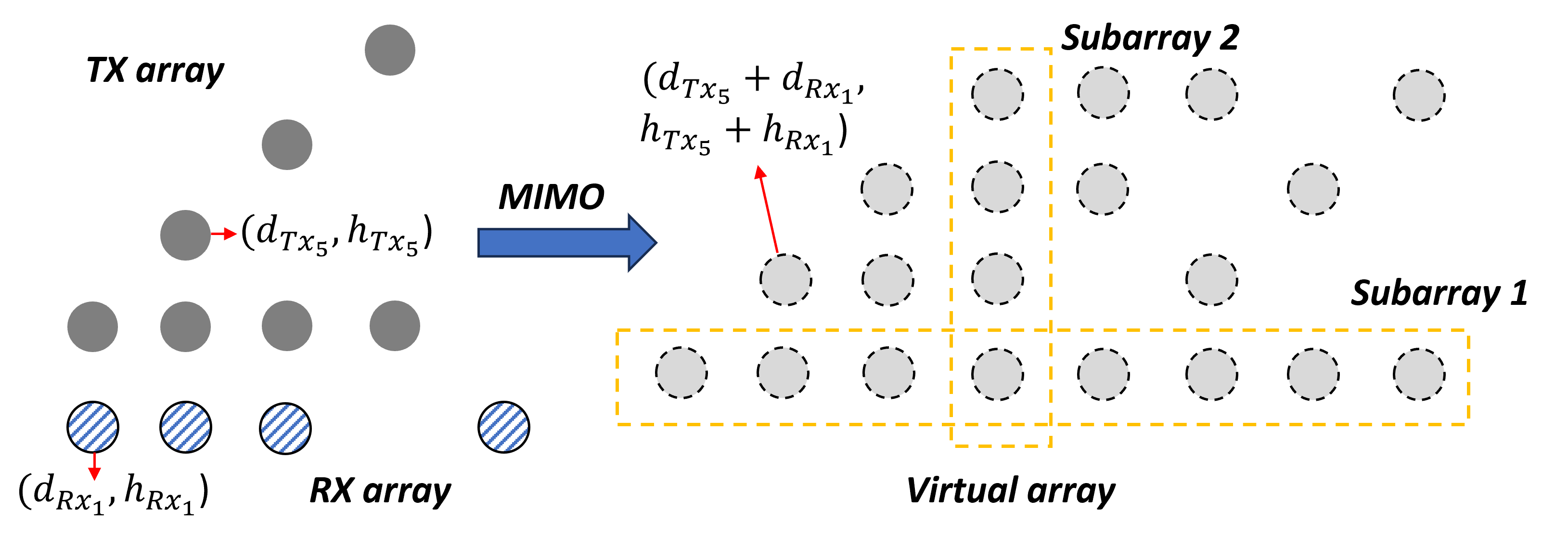}
\caption{A non-uniform MIMO array with $7$ TXs and $4$ RXs} and the resulting $28$-element virtual array (some virtual elements are overlapped). The location of the 1st RX and 5th TX elements and the resulting virtual element are labeled.
  \label{fig:virtual}
\end{figure}

Here, we consider a general TX array with \(P\) TX elements and a receiver (RX) array with \(Q\) RX elements. The horizontal and vertical locations of TX elements and RX elements are denoted by \((d_{\mathrm{Tx}_p}, h_{\mathrm{Tx}_p})\) and \((d_{\mathrm{Rx}_q}, h_{\mathrm{Rx}_q})\), respectively, where \(p \in \{1, 2, \dots, P\}\) and \(q \in \{1, 2, \dots, Q\}\). The spacing between TX elements and RX elements can be either uniform or non-uniform, as illustrated in Fig.~\ref{fig:virtual}. The orthogonality of the transmit signals can be ensured through MIMO techniques, enabling the recovery of individual transmitted signals at the RX array. The RX array can be extended to a larger ``virtual array" by stacking the measurements at the physical receive array corresponding to each orthogonal TX waveform \cite{ti_mimo}. The location of the MIMO virtual arrays is the spatial convolution of the TX and RX arrays, resulting in \(PQ\) virtual elements in total, with each virtual element being a product of a TX-RX element pair. For instance, the virtual array location corresponding to the \(p\)-th TX element and \(q\)-th RX element is \((d_{\mathrm{Tx}_p} + d_{\mathrm{Rx}_q}, h_{\mathrm{Tx}_p} + h_{\mathrm{Rx}_q})\).

There are several methods to achieve TX signal orthogonality, including frequency division multiplexing, time division multiplexing, code division multiplexing, and Doppler division multiplexing, among others. In this paper, we focus on the most popular approach, time division multiplexing (TDM), for the MIMO setup and model its radar signal. The TDM-MIMO scheme ensures the orthogonality of transmit signals by sequentially transmitting chirps from each TX.

From array theory \cite{gao2021perception}, for a far-field target with azimuth angle \(\theta\) and elevation angle \(\varphi\), the signal for it at the virtual antenna with location \((d_{\mathrm{Tx}_p} + d_{\mathrm{Rx}_q}, h_{\mathrm{Tx}_p} + h_{\mathrm{Rx}_q})\) can be modeled by the phase terms induced by the target's azimuth and elevation angles: \(\frac{2\pi}{ \lambda} \left((d_{\mathrm{Tx}_p} + d_{\mathrm{Rx}_q})\cos{\theta}+(h_{\mathrm{Tx}_p} + h_{\mathrm{Rx}_q})\sin{\varphi}\right)\). In the TDM-MIMO setup, the interval of transmitting one chirp by \(P\) TXs is extended to \(P T_c\) (which differs from the chirp duration \(T_c\) in the single TX case), resulting in a new round-trip delay \(\tau_{\mathrm{TDM}, n}\) for the target during the \(n\)-th chirp:
\begin{equation}
    \tau_{\mathrm{TDM}, n} = \tau_{0} - \frac{2 v_{\R{r}} n P T_{\R{c}}}{c_0} = =  \tau_{0} + n \Delta \tau_{\mathrm{TDM}},
\label{eq:tau_new}
\end{equation} 
\noindent where \(\Delta \tau_{\mathrm{TDM}}=-\frac{2 v_{\mathrm{r}} P T_{\mathrm{c}}}{c_0}\). If a relative motion between the radar platform and the target is present, there will be an additional Doppler phase term for different TXs as they transmit the signal at different times. In general, for \(P\) TXs the phase relation at the \(p\)-th TX is given by \cite{8052088}:
\begin{equation}
\Delta \tilde{\varphi}_{\mathrm{Tx}_p}= j 2 \pi f_{\R{c}} \Delta \tau_{\mathrm{TDM}} \frac{p-1}{P}= - j 4 \pi (p-1) \frac{v_{\R{r}} T_{\R{c}}}{\lambda}.
\end{equation} 

Therefore, the IF signal for chirp \(n\), fast-time sample \(m\), and \(p\)-th TX and \(q\)-th RX element in TDM-MIMO can be updated from Eq.~\eqref{eq:ifs_dop} and given by: 
\begin{dmath}
    s_{\R{IF}}(m, n, p, q)=\\
    \frac{A_{\R{T}} A_{\R{R}}\beta}{2} \exp \left( j2 \pi\left(S_{\R{w}} \tau_{\mathrm{TDM}, n} \frac{m}{f_{\R{s}}} + f_{\R{c}} \tau_{\mathrm{TDM}, n} - \frac{1}{2} S_{\R{w}} {\tau_{\mathrm{TDM}, n}}^{2} + \frac{(d_{\mathrm{Tx}_p} + d_{\mathrm{Rx}_q})\cos{\theta}}{\lambda} + \frac{(h_{\mathrm{Tx}_p} + h_{\mathrm{Rx}_q})\sin{\varphi}}{\lambda} \right) + \Delta \tilde{\varphi}_{\mathrm{Tx}_p} \right).
 \label{eq:sifmimo}
\end{dmath}

\begin{figure*}[]
\centering
\includegraphics[width=0.89\textwidth]{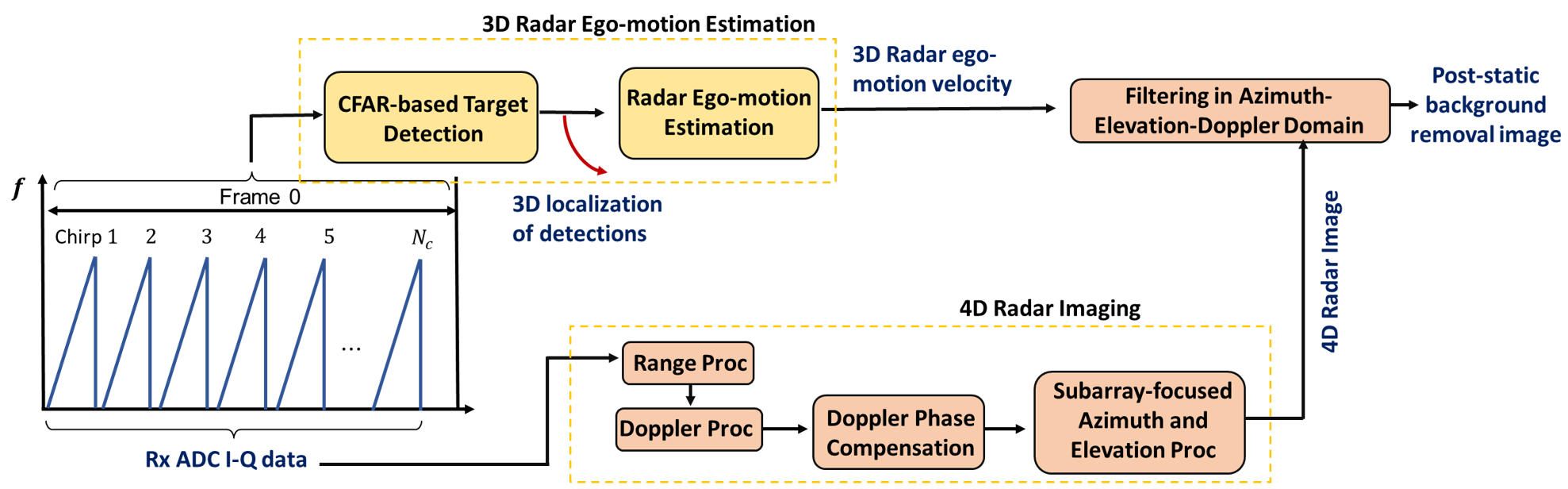}
\caption{The overall workflow for the proposed static background removal algorithm is depicted by the red blocks, comprising a 4D radar imaging step and filtering in the azimuth-elevation-Doppler domain. Additionally, the filtering process necessitates a known 3D radar ego-motion velocity, obtainable either through an external sensor or the proposed self-contained 3D ego-motion estimation approach illustrated in the yellow blocks.}
  \label{fig:sys_workflow}
\end{figure*}

\section{Static Background Removal Algorithm}

The proposed static background removal algorithm for automotive radars operates in a frame-by-frame manner, as illustrated in the red blocks of Fig.~\ref{fig:sys_workflow}. It begins by processing one frame of radar post-ADC data, $s_{\R{IF}}(m, n, p, q)$, to generate a 4D radar image (range-Doppler-azimuth-elevation). Using the known radar ego velocity $(v_{x},v_{y},v_{z})$, the algorithm identifies Doppler frequencies corresponding to the static background based on the specific Doppler velocity profile. It then removes the background via notch filtering in the azimuth-elevation-Doppler domain, effectively isolating moving targets.

\subsection{4D Radar Imaging}
\label{sec:4d_image}
The 4D radar imaging process is composed of four steps: range processing, Doppler processing, Doppler phase compensation, and subarray-focused azimuth and elevation processing.

\subsubsection{Range Processing}
The input IF signal has a beat frequency \(f_{\mathrm{b}}=S_{\mathrm{w}} \tau_{\mathrm{TDM}, n}\), where \(\tau_{\mathrm{TDM}, n}\) represents the round-trip delay corresponding to the target's distance. To estimate this beat frequency, a fast Fourier transform (FFT), referred to as the {\em Range FFT}, is applied to convert the time-domain IF signal into the frequency domain \cite{gao2019experiments}. Peaks in the resulting spectrum (or range profile) are then used to estimate the target's distance.

Mathematically, the Range FFT implemented on chirp \(n\), TX \(p\), and RX \(q\) is given by:
\begin{equation}
    S_{\mathrm{R}}(m_{\mathrm{r}}, n, p, q) = \mathcal{F}_m\{ s_{\mathrm{IF}}(m, n, p, q)\},
\end{equation}
where \(m_{\mathrm{r}}\) represents the range bin index and \(\mathcal{F}\) is the FFT operation. The range resolution is determined by the swept RF bandwidth \(B\) with the well-known equation \(R_{\mathrm{res}}=\frac{c_0}{2B}\).

\subsubsection{Doppler Processing}
According to Eq.~\eqref{eq:tau_new} and \eqref{eq:sifmimo}, the relative radial velocity \(v_{\mathrm{r}}\) will cause a Doppler phase shift \(\Delta \phi_{\mathrm{v}}= -\frac{4\pi v_{\mathrm{r}} P T_{\mathrm{c}}}{\lambda}\) in the IF signal between consecutive chirps. Hence, a fast Fourier transform (referred to as {\em Doppler FFT}) is executed across chirps to estimate the phase shift and then transform it to velocity \cite{gao2019experiments}.

Mathematically, the Doppler FFT performed on the range profile is expressed as:
\begin{equation}
    S_{\mathrm{RV}}(m_{\mathrm{r}}, n_{\mathrm{v}}, p, q) = \mathcal{F}_n\{ S_{\mathrm{R}}(m_{\mathrm{r}}, n, p, q) \},
\end{equation}
where \(n_{\mathrm{v}}\) is the velocity bin index. The velocity resolution of this method is given by \(V_{\mathrm{res}}=\frac{\lambda}{2N_{\mathrm{c}} PT_{\mathrm{c}}}\) \cite{iovescu2017fundamentals}, where \(N_{\mathrm{c}}\) is the number of chirps in a frame.

\subsubsection{Doppler Phase Compensation}
Under the TDM-MIMO setup, motion-induced phase errors for non-stationary targets should be compensated before making azimuth and elevation estimations \cite{gao2019experiments, gao2021perception}. According to \cite{8052088}, these errors are corrected via phase compensation of \(\Delta \tilde{\varphi}_{\mathrm{Tx}_p}\), which can be obtained from the Doppler FFT results. We denote the phase compensation term of Doppler bin \(n_{\mathrm{v}}\) and TX \(p\) by \(\Delta \tilde{\varphi}_{\mathrm{Tx}_p}(n_{\mathrm{v}})\). We can mathematically model the Doppler phase compensation step as:
\begin{equation}
    S_{\mathrm{RV}_{\text{comp}}}(m_{\mathrm{r}}, n_{\mathrm{v}}, p, q) = S_{\mathrm{RV}}(m_{\mathrm{r}}, n_{\mathrm{v}}, p, q) \cdot \Delta \tilde{\varphi}_{\mathrm{Tx}_p}(n_{\mathrm{v}}).
\end{equation}

\begin{figure}
\centering
\includegraphics[width=0.49\textwidth]{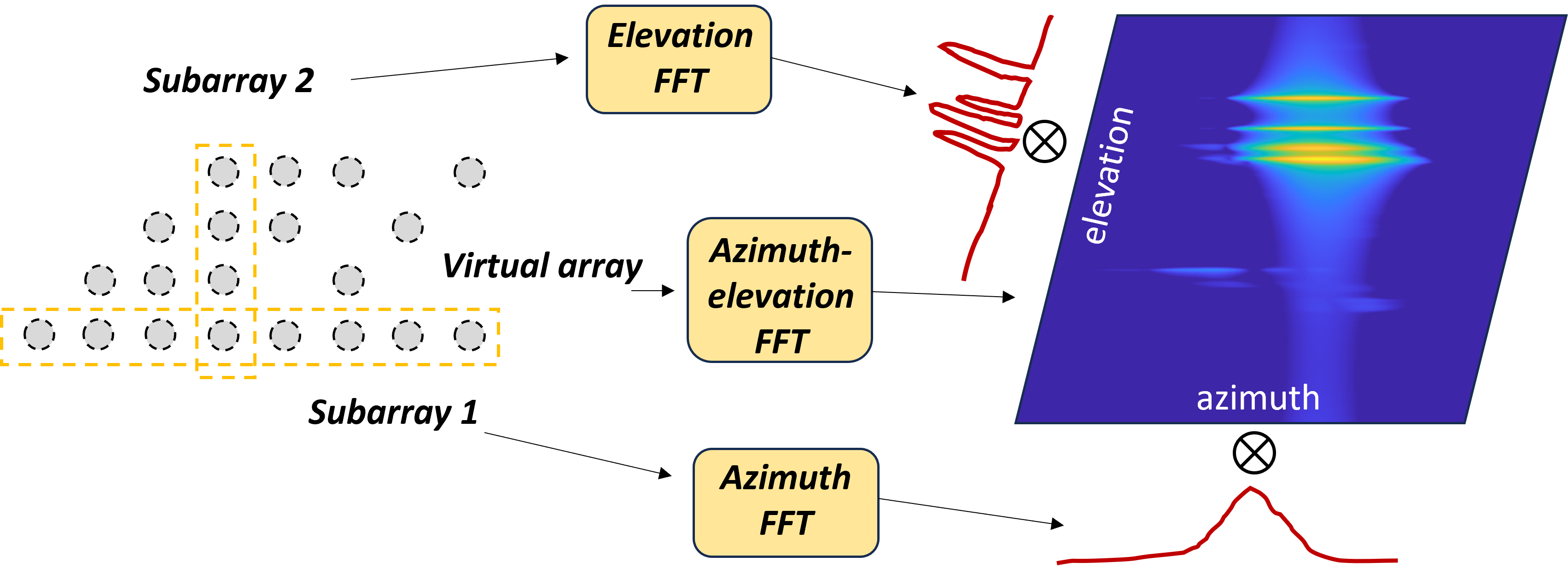}
\caption{The workflow of subarray-focused azimuth and elevation processing. The spectrums of horizontal and elevation subarrays are multiplied with the Azimuth-Elevation FFT results of a 2D non-uniform virtual array to effectively suppress the sidelobes.}
  \label{fig:subarray_aoa}
\end{figure}

\subsubsection{Subarray-focused Azimuth and Elevation Processing}
For a 2-dimensional (2D) non-uniform array, although resolution benefits from increased aperture, azimuth and elevation estimation results suffer from high sidelobes due to empty elements in the non-uniform array \cite{gao2021perception}. To address this issue, we propose a subarray-focused azimuth and elevation processing approach. The motivation behind this approach is that within the non-uniform array, there are subarrays with larger apertures and fewer missing elements, which can be used to classify mainlobes and sidelobes, thereby suppressing the sidelobes.

In this approach, the first step is to select from the virtual array (Fig.~\ref{fig:subarray_aoa}) a horizontal subarray (subarray 1) and an elevation subarray (subarray 2) with the largest aperture and the least missing elements. Then, for each range-Doppler bin of \(S_{\mathrm{{RV}_{comp}}}(m_{\mathrm{r}}, n_{\mathrm{v}}, p, q)\), we perform a fast Fourier transform on the selected horizontal subarray (referred to as {\em Azimuth FFT}) and a fast Fourier transform on the selected elevation subarray (referred to as {\em Elevation FFT}) to obtain an azimuth spectrum and an elevation spectrum. Subsequently, the two spectrums are normalized to a maximum value of 1 for sidelobe suppression in the next step.

The second step involves making a joint azimuth and elevation estimation by employing a 2D fast Fourier transform on the entire virtual array (referred to as {\em Azimuth-Elevation FFT}). The missing elements are filled with zero values for the input to the FFT. Mathematically, the azimuth-elevation FFT is represented as:
\begin{equation}
    S_{\mathrm{RVAE}}(m_{\mathrm{r}}, n_{\mathrm{v}}, p_\theta, q_\varphi) =  \mathcal{F}_q\{\mathcal{F}_p\{ S_{\mathrm{{RV}_{comp}}}(m_{\mathrm{r}}, n_{\mathrm{v}}, p, q) \} \},
\end{equation}
where \(p_\theta\) is the azimuth angle bin and \(q_\varphi\) is the elevation angle bin. It is expected that a non-uniform 2D virtual array will introduce a large number of sidelobes in the azimuth-elevation FFT results \(S_{\mathrm{RVAE}}(m_{\mathrm{r}}, n_{\mathrm{v}}, p_\theta, q_\varphi)\). Next, as shown in Fig.~\ref{fig:subarray_aoa}, we reduce the unnecessary sideslobes on \(S_{\mathrm{RVAE}}\) by multiply \(S_{\mathrm{RVAE}}\) by the normalized azimuth spectrum and elevation spectrum from the subarrays in step 1. The sidelobes are attenuated because the subarray spectra yield low values after normalization in the non-mainlobe areas. The final output \(S_{\mathrm{RVAE}}\) is the 4D radar image depicting the rage-Doppler-azimuth-elevation of the environment.

\subsection{Filtering in Azimuth-Elevation-Doppler Domain}
\label{sec:removal}

With a known radar ego-motion velocity \(\mathbf{v}_{\mathrm{c}}\) and 4D radar images generated from the steps in the previous section, we exploit the relationship among radar velocity, target velocity, and target azimuth and elevation angles to eliminate the reflection of the static background from the radar images. Specifically, we calculate the expected Doppler velocity \(v_{\mathrm{r}}\) for each feasible azimuth-elevation angle pair \((\theta, \varphi)\) that satisfies the relationship in Eq.~\eqref{eq:vr}. Subsequently, we employ a 3D notch filter to remove the Doppler component for specific azimuth and elevation from the radar image.

Notch filtering involves the removal or suppression of specific components in the Doppler spectrum by element-wise multiplication with the frequency response of the filter. The frequency response of a second-order one-dimensional (1D) infinite impulse response (IIR) digital notch filter is defined by \cite{wang2013second} as:
\begin{equation}
H(z)=\bigg| \frac{1-2 \cos \omega_0 z^{-1}+z^{-2}}{1-2 s \cos \omega_0 z^{-1}+s^2 z^{-2}} \bigg|,
\label{eq:notch_1d}
\end{equation}
\noindent where \(\omega_0\) represents the notch filter center frequency, \(z=e^{j\omega}\), and \(s\) is a coefficient satisfying \(0\leq r < 1\).

In our scenario, a 3D notch filter is required to filter out components in the joint Doppler-azimuth-elevation dimension. The construction of a 3D notch filter involves obtaining its frequency response by combining the frequency responses of three 1D notch filters. The specific steps are outlined below. First, design three separate 1D notch filters following Eq.~\eqref{eq:notch_1d} to reject frequencies corresponding to the azimuth angle \(\theta\), elevation angle \(\varphi\), and Doppler \(v_{\mathrm{r}}\), respectively. Second, expand the frequency response of each 1D filter to 3D by replicating it along the other two dimensions \cite{ramp}. Third, take the point-wise minimum among the three expanded 3D filters to construct the frequency response for the desired 3D notch filter. This methodology offers a simplified yet effective approach for designing the 3D notch filter.

To remove the static background from the 4D radar image, we perform element-wise multiplication between the radar image \(S_{\mathrm{RVAE}}\) and the frequency response of the 3D notch filter, denoted as \(|H_{3D}(z_{\theta_i}, z_{\varphi_i}, z_{v_{r_i}})|\), constructed above for each angle pair \((\theta_i, \varphi_i)\) and corresponding Doppler velocity \(v_{r_i}\). Assuming there are a total of \(M\) angle pairs covering all azimuth and elevation angles, the background-removed radar image \(S_{\mathrm{remove}}\) can be calculated as follows:
\begin{equation}
S_{\mathrm{remove}} = S_{\mathrm{RVAE}} \otimes \prod_{i=1}^{M}\big|H_{3D}(z_{\theta_i}, z_{\varphi_i}, z_{v_{r_i}})\big|,
\end{equation}
where \(\otimes\) represents the element-wise multiplication.

The computational burden associated with constructing the notch filter can be notably mitigated by employing a fixed 3D notch filter that can be readily adapted to different notch frequencies \((\theta_i,\varphi_i, v_{r,i})\) through simple frequency translation. This obviates the necessity of designing a new filter for each specific notch frequency. Additionally, the frequency response of the notch filter can be confined to a narrow region around the desired notch frequency \((\theta_i,\varphi_i, v_{r,i})\). By concentrating solely on a limited area, the number of operations required for the element-wise multiplication between the filter and the radar image can also be substantially reduced.


\section{3D Radar Ego-motion Estimation} 
\label{sec:odr}
The traditional solution to obtaining 3D radar ego-motion is to use on-board inertial measurement units (IMU) \cite{kong2000inertial, mimosar}, which combine measurements from the wheel speed sensor, gyroscopes, and accelerometers. However, high-precision IMUs are cost-prohibitive for automotive applications, which inspires the need for self-contained alternatives such as radar odometry to determine the velocity and direction of motion of the vehicular radar \cite{mimosar, 9096265, 8995552}. In this section, we introduce an approach to estimate radar ego-motion from the radar itself by extracting radar targets and analyzing the distribution of the radial velocities.

\subsection{CFAR-based Target Detection}
\label{sec:point}
\begin{figure}
\centering
\includegraphics[width=0.49\textwidth]{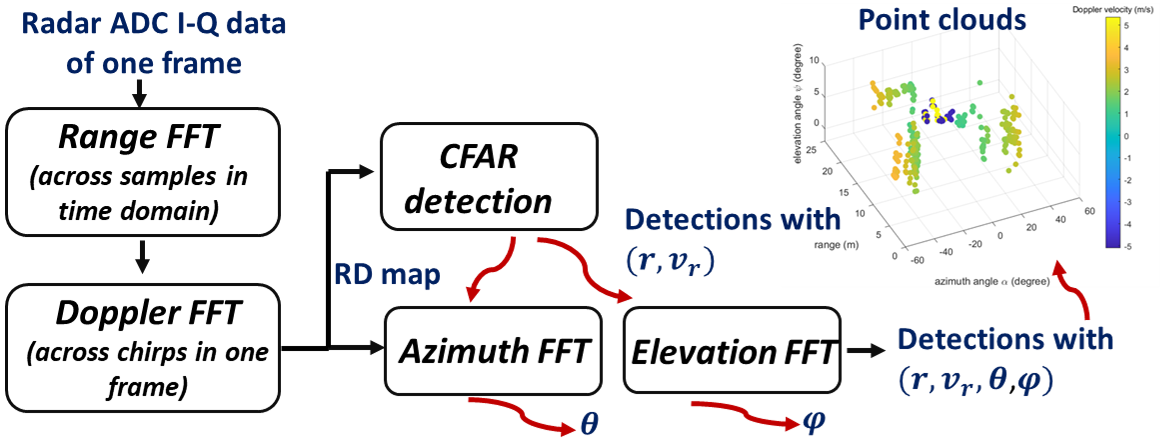}
\caption{Workflow for processing radar ADC data to extract radar detections with corresponding $(r, v_{\R{r}}, \theta, \varphi)$ values.}
  \label{fig:point}
\end{figure}

The point cloud extraction process combines basic FFT processing, as discussed in Section~\ref{sec:4d_image}, with cell-averaging constant-false-alarm-rate (CA-CFAR) detection techniques \cite{gao2019experiments}. The workflow, depicted in Fig.~\ref{fig:point}, takes ADC I-Q data as input and yields a collection of detections represented by their \((r, v_{\mathrm{r}}, \theta, \varphi)\) values, forming `point clouds'.

The initial step involves estimating range and Doppler velocity through two sequential FFTs: the Range FFT and Doppler FFT \cite{gao2019experiments}. The resulting Range-Doppler (RD) map undergoes processing by the CA-CFAR algorithm \cite{piofmd} to detect peaks and derive their corresponding ranges and Doppler velocities \((r, v_{\mathrm{r}})\). During CA-CFAR detection, each cell in the RD map is assessed for the presence or absence of a target using a threshold, determined based on the noise power estimate within the training cells. For the RD spectrum cell corresponding to \((r, v_{\mathrm{r}})\), Azimuth FFT and Elevation FFT processing, as described in Section~\ref{sec:4d_image}, are employed to estimate \((\theta, \varphi)\) values using a horizontal or vertical subarray (in the virtual array). In the case of non-stationary targets, motion-induced phase errors at each virtual antenna location due to TDM-MIMO should be compensated for different TXs before angle estimation \cite{8052088}.

\subsection{Radar Ego-motion Estimation}
We assume that there are \(N_{\mathrm{total}}\) detections (indexed by \(i\)) in the extracted point clouds, identified by \((r_i, v_{r,i},\theta_i, \varphi_i)\), a subset of which belongs to the class of stationary targets, initially unknown. Representing the size of the stationary subset by \(N\), from Eq.~\eqref{eq:vr}, the relationship between the Doppler velocities and angles of \(N\) stationary targets (\(i=1,2,\dots, N\)) can be expressed as follows:

\begin{equation}
\arraycolsep=1.4pt
\resizebox{.98\hsize}{!}{
    $\left[\begin{matrix} v_{r,1}\\ \vdots\\v_{r,N}\\
    \end{matrix}\right] =-\left[\begin{matrix} \sin{\theta_1}\cos{\varphi_1}&\cos{\theta_1}\cos{\varphi_1}&\sin{\varphi_1} \\ \vdots&\vdots&\vdots\\ \sin{\theta_N}\cos{\varphi_N}&\cos{\theta_N}\cos{\varphi_N}&\sin{\varphi_N}\\
    \end{matrix}\right] \left[\begin{matrix}
    v_x\\v_y\\v_z
    \end{matrix}\right]  + 2kv_{\R{max}}
    $}.
    \label{eq:odo_mat}
\end{equation}

Here, \(v_{\mathrm{max}}\) is the radar's maximum unambiguously measurable Doppler velocity, \(k \in \mathbb{Z}\) is an integer describing Doppler ambiguity, assuming all Doppler measurements from static subsets are aliased with the same \(k\) value. The current trend in 4D imaging radar \cite{ti_casd} is towards increasing the density of MIMO arrays. Hence, as the number of TXs \(N_T\) increases, it leads to a proportional reduction in the maximum unambiguously measurable velocity \(v_{\mathrm{max}}\) \cite{gao2019experiments}, following \(v_{\mathrm{max}} = \frac{\lambda}{4N_T T_c}\), causing Doppler ambiguity whenever the true Doppler exceeds \(v_{\mathrm{max}}\).

If \(k\) is a known value, we can rearrange the \(2kv_{\mathrm{max}}\) term to the left side of Eq.~\eqref{eq:odo_mat} and utilize the least squares regression (LSR) to solve \((v_x, v_y, v_z)\) \cite{mimosar}. However, in reality, scenarios involve moving objects, resulting in a mixture of stationary and moving object detections obtained from Section~\ref{sec:point}, which do not conform to the model in Eq.~\eqref{eq:odo_mat}. Hence, we employ the Random Sample Consensus (RANSAC) algorithm \cite{10.1145/358669.358692} along with LSR to separate the required stationary targets and determine \(N_{\mathrm{d}}\) \cite{mimosar,6728341}. RANSAC is an iterative method for optimally extracting inliers (corresponding to stationary targets) that fit the model Eq.~\eqref{eq:odo_mat} well and separating outliers (moving targets or clutter) by randomly sampling observed data \cite{10.1145/358669.358692}.

While \(k\) is indeed unknown, bounds on possible \(k\) values in a set \({\cal K}\) can be readily determined in the real world, depending on \(v_{\mathrm{max}}\) and the maximum vehicle driving speed. For instance, if \(v_{\mathrm{max}} = \SI{5}{m/s}\) and the maximum vehicle driving speed is \(\SI{11.2}{m/s}\) (the speed limit for urban street driving), then \(k \in {\cal K} = \{ -1, 0, \text{or}\, 1 \}\). This is because when \(k\) is -1 or 1, the maximum/minimum Doppler velocity of any stationary target can be folded into the range of \([-v_{\mathrm{max}}, v_{\mathrm{max}}]\) by adding \(2kv_{\mathrm{max}}\).

Based on the above analysis, we propose a heuristic approach to solve Eq.~\eqref{eq:odo_mat}. First, we determine a set \({\cal K}\) to bound possible \(k\). Then, for each \(k \in \cal K\), we substitute it into Eq.~\eqref{eq:odo_mat} and apply least squares to solve \((v_x, v_y, v_z)\). After iterating through all \(k\), we can determine the optimal \(k^*\) that yields the maximum number of inliers (from the RANSAC output). In other words, we choose the \(k\) value that maximizes the number of data points that satisfy Eq.~\eqref{eq:odo_mat}. Consequently, the best-fit radar ego velocity \(\mathbf{v}_{\mathrm{c}}^*=(v^*_x, v^*_y, v^*_z)\) would be the one corresponding to the optimal \(k^*\).

\begin{figure}
\centering
\includegraphics[width=0.45\textwidth]{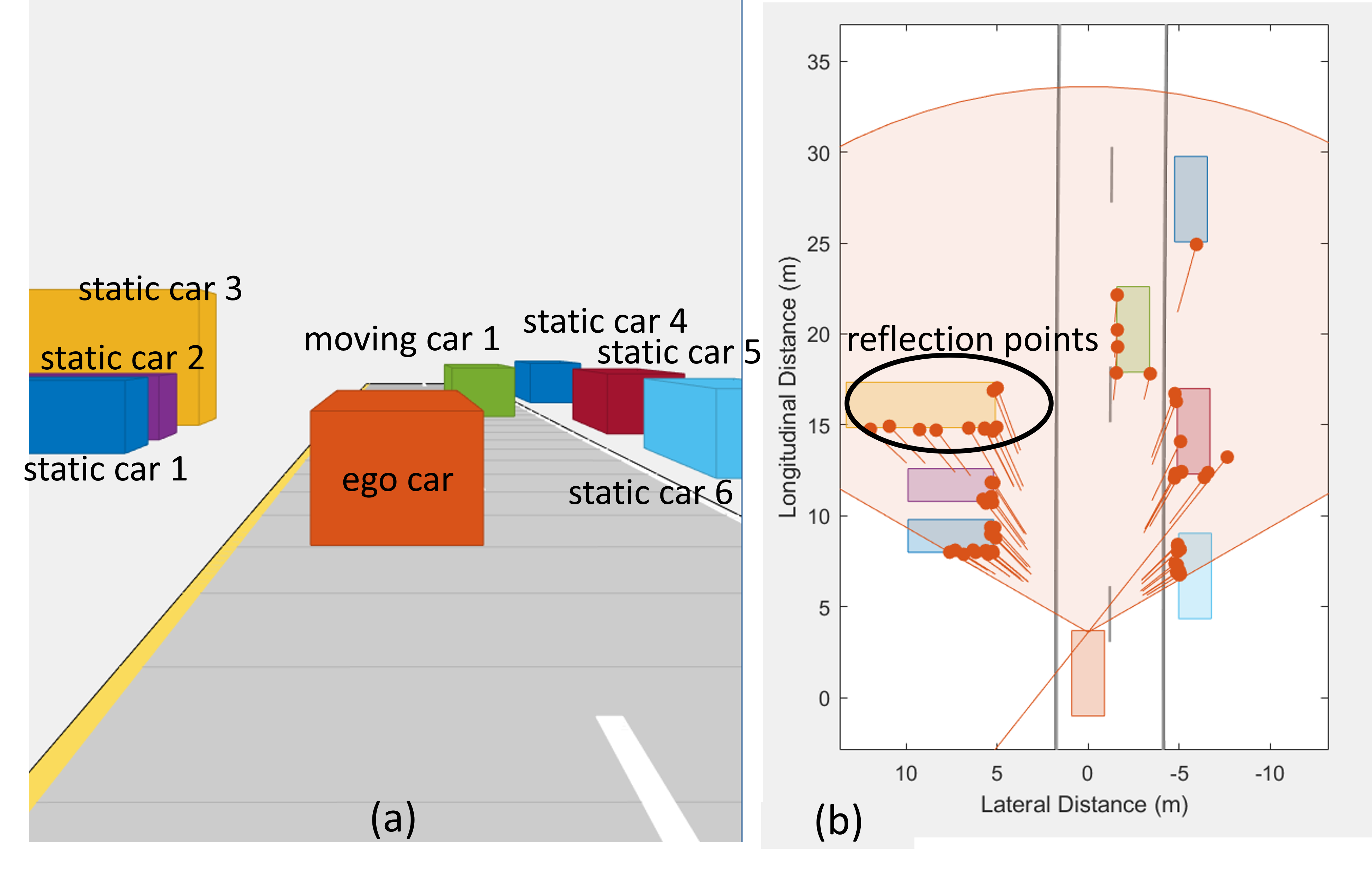}
\caption{(a) Simulation scenario on MATLAB. (b) The simulated reflection points for targets presented in the bird's-eye view.}
  \label{fig:simu}
\end{figure}

\begin{figure}
\centering
\includegraphics[width=0.48\textwidth]{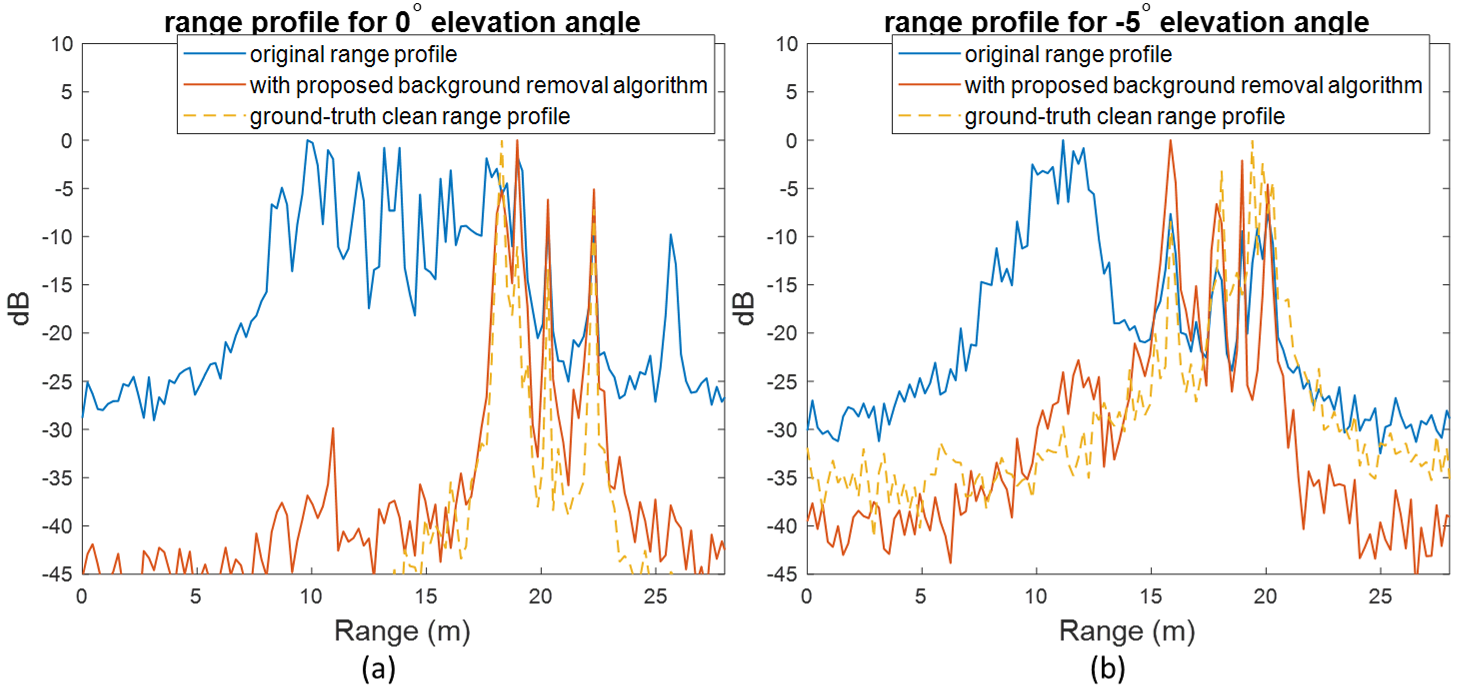}
\caption{Simulated static background removal results on range profile results for different elevation angles.}
\label{fig:rng_profile}
\end{figure}

\begin{figure*}
\centering
\includegraphics[width=0.95\textwidth]{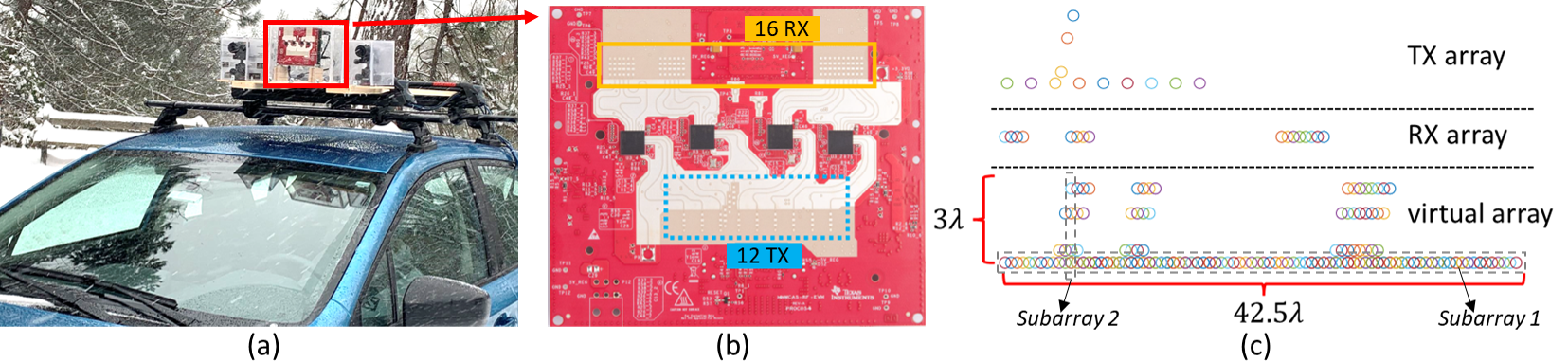}
\caption{(a) Experimental radar testbed mounted on a vehicle. (b) The front view of the cascaded-chip radar board with the RX and TX highlighted. (c) The arrangement for the TX array, RX array, the formed virtual array using TDM-MIMO, and the selected subarrays for the subarray-focused azimuth and elevation processing.}
\label{fig:testbed}
\end{figure*}

\section{Simulations}
\subsection{Simulated Scenario and Configuration \label{sec_simu_scenario}}
\subsubsection{Scenario}
We simulated a typical driving scenario using the MATLAB Automated Driving Toolbox, involving an ego-moving vehicle along a linear road, three parked vehicles each on either side of the road, and one moving vehicle ahead of the ego-moving vehicle, as illustrated in Fig.~\ref{fig:simu}(a). The ego-motion car has a forward velocity of $(\SI{8}{m/s}, 0, \SI{-0.5}{m/s})$ and an acceleration of $(\SI{2}{m/s^2}, \SI{1}{m/s^2}, 0)$ and is equipped with a front-view FMCW radar. The other moving car had a constant velocity of $(\SI{5}{m/s}, 0, 0)$. The parked cars were represented by a collection of 3D reflection points within the radar's field of view, as indicated by the red markers in Fig.~\ref{fig:simu}(b). Using the reflection points and the signal model described by Eq.~\eqref{eq:sifmimo}, we generated post-demodulated ADC samples for each frame at the receiver, considering the specific radar configuration described next.

\subsubsection{Configuration}
The radar configurations used for generating the ADC samples are as follows: $f_{\R{c}}=\SI{77}{GHz}$, $S_{\R{w}}=\SI{21.0017}{MHz/us}$, $A_{\R{T}}=A_{\R{R}}=1$, $\phi_{0}=0$, $f_{\R{s}}=\SI{4}{Msps}$, the number of samples per chirp is 128, the number of chirps per frame sent by each TX is 255. We assume the formed virtual array is  planar with $8 \times 8$ elements and antenna distance $h=\lambda/2$, where $\lambda$ is the wavelength $c_0/f_{\R{c}}$. We consider the pulse repetition interval for TX to be \SI{60}{us}, corresponding to a maximum unambiguous Doppler velocity of $\SI{16.5}{m/s}$, which creates no Doppler ambiguity for most of the city/town street driving cases. For evaluation purposes, we simulated a total of 40 frames with a frame rate of $\SI{20}{fps}$. 

\subsection{Implementation and Analysis} 
We applied the proposed 3D radar ego-motion estimation algorithm and static background removal algorithm to the simulated data. To quantitatively assess the performance of the background removal algorithm, we analyzed the range profiles before and after the background removal process, as shown in Fig.~\ref{fig:rng_profile}, and compared them with the simulated ground-truth clean signal. The range profiles are plotted with the average amplitudes (in dB) for different elevation angles, as illustrated in Fig.~\ref{fig:rng_profile}. For an elevation angle of \SI{0}{\degree}, before background removal, the original signal amplitude (blue) at \SI{8}{m} to \SI{18}{m} was around \SI{-10}{dB}, while the interference amplitude was \SI{-25}{dB}, resulting in a signal-to-interference ratio (SIR) of \SI{15}{dB}. After background removal, the signal amplitude increased to \SI{0}{dB}, and the interference amplitude significantly decreased to \SI{-40}{dB}, yielding a greatly improved SIR of \SI{40}{dB}. This substantial improvement in SIR from \SI{15}{dB} to \SI{40}{dB} demonstrates the remarkable clutter suppression capability of the proposed algorithm. Additionally, both Fig.~\ref{fig:rng_profile}(a) and (b) show that the post-background removal curve aligns well with the ground-truth clean range profile, indicating the accuracy of the proposed algorithm.

\section{Experiments}
Besides simulations, we also gathered significant measurement data using an off-the-shelf automotive radar testbed. This testbed allows us to assess the performance of the proposed static background removal algorithm in a real-world setting.

\subsection{Experimental Setup and Configuration}
\subsubsection{Setup}
A testbed was constructed using Texas Instruments's TIDEP-01012 radar \cite{ti_casd} and two cameras, as depicted in Fig.~\ref{fig:testbed}(a). The radar evaluation module operates in the \SI{77}{GHz}-\SI{81}{GHz} band and employs a cascade of four radar chips, enabling a greater MIMO dimension \cite{ti_casd}. For data capture, the testbed was positioned at the front of the vehicle, enabling simultaneous collection of camera images and corresponding radar raw ADC data \cite{ramp}. The front view of the radar setup is illustrated in Fig.~\ref{fig:testbed}(b), featuring a 2D arrangement with 12 TXs and 16 RXs. By employing TDM MIMO techniques \cite{gao2019experiments, gao2021perception}, the orthogonal signals from different TXs result in the formation of a 2D virtual receiver array, as shown in Fig.~\ref{fig:testbed}(c), obtained through the spatial convolution of all physical TX and RX pairs \cite{gao2019experiments, gao2021perception}. The virtual array exhibits a sparse configuration (with 4 non-uniformly spaced elements spanning a $3\lambda$ aperture) in the vertical direction, and a large uniform array (consisting of 86 uniformly spaced elements spanning a $42.5\lambda$ aperture) in the horizontal direction.

\begin{figure}
\centering
\includegraphics[width=0.4\textwidth]{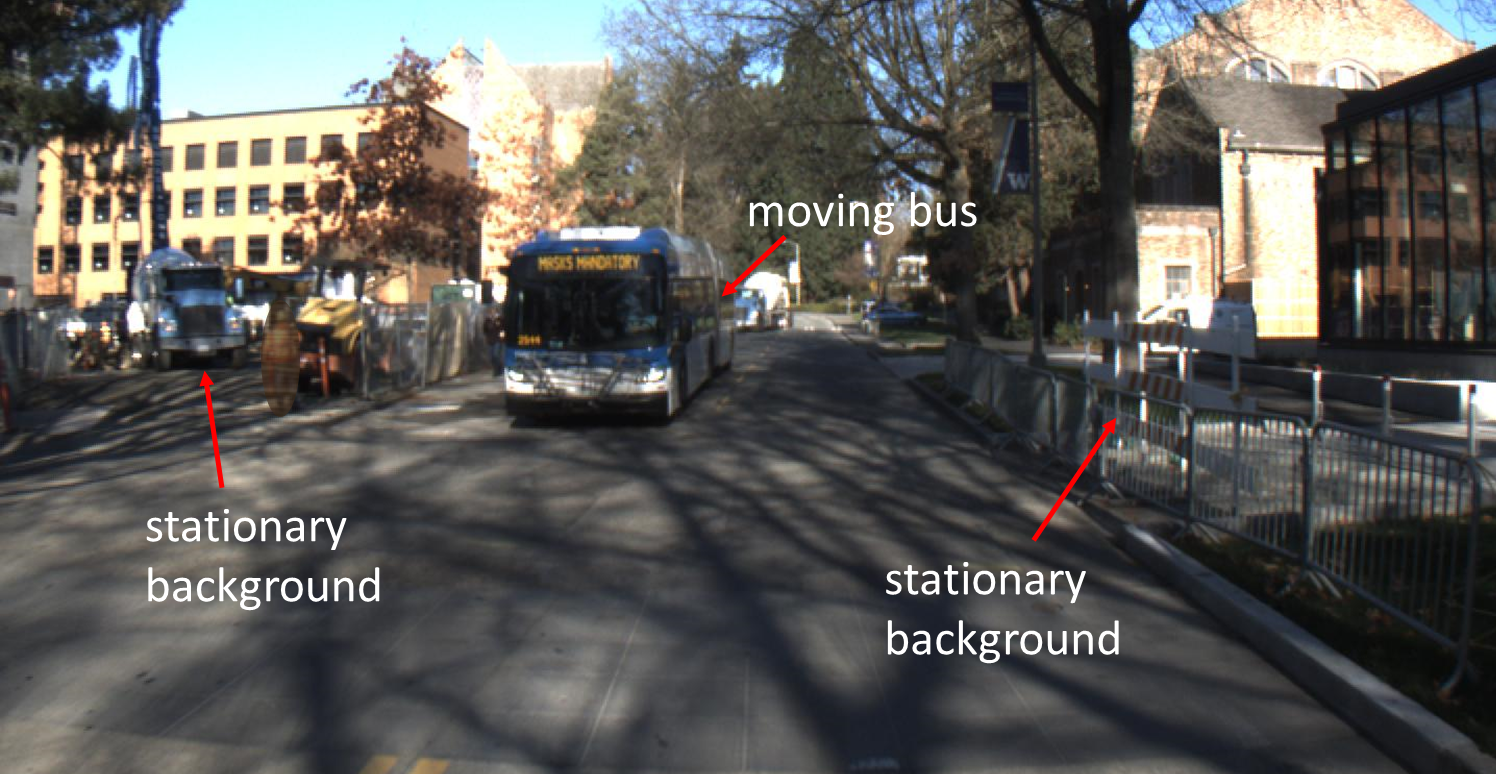}
\caption{Visualization of the experiment scene.}
  \label{fig:imag_exp}
\end{figure}

\begin{figure*}
\centering
\includegraphics[width=0.98\textwidth]{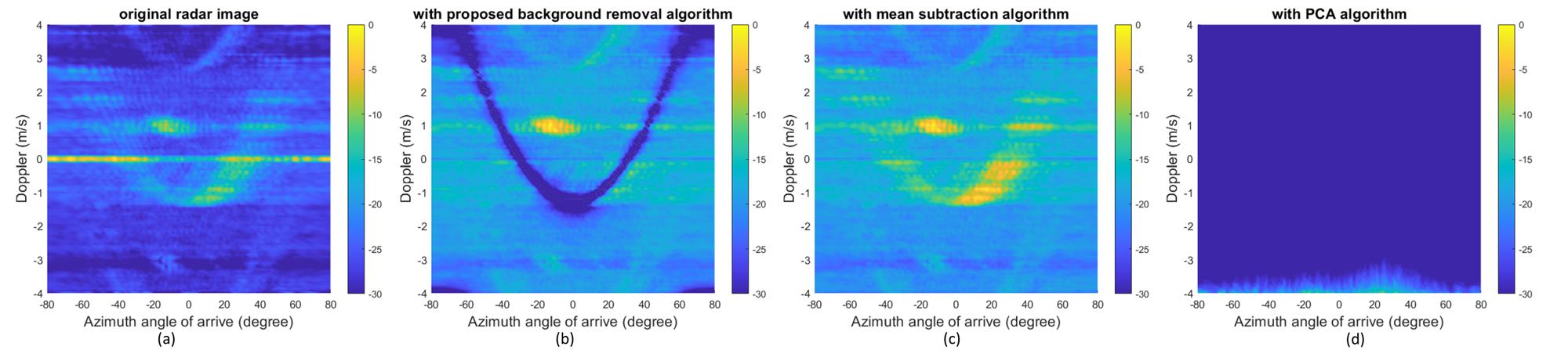}
\caption{Doppler-azimuth angle images (at \SI{0}{\degree} elevation angle) for different algorithms: (a) original radar image, (b) proposed background removal algorithm, (c) mean subtraction algorithm, and (d) PCA algorithm.}
 \label{fig:dop_agl_maps_exp}
 \includegraphics[width=0.98\textwidth]{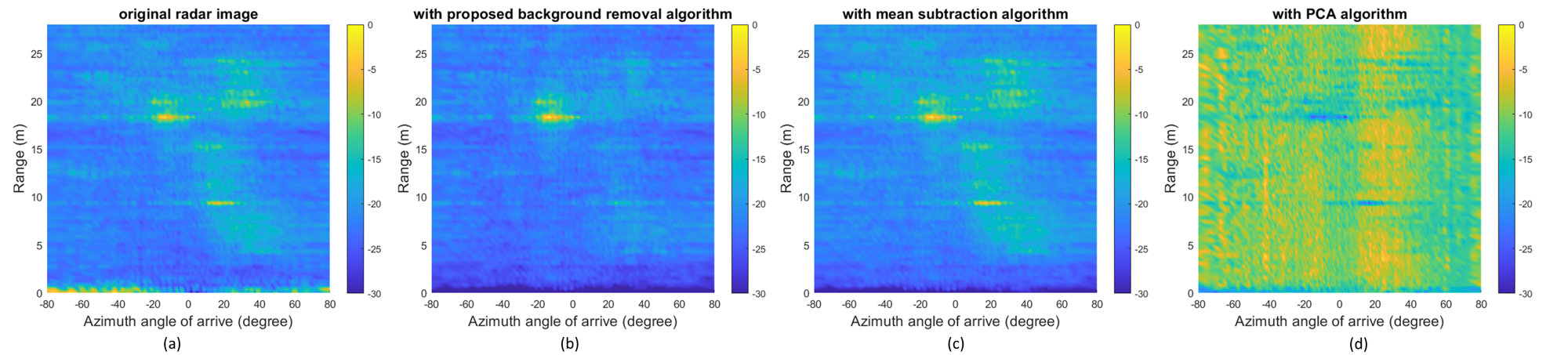}
\caption{Range-azimuth angle images (at \SI{0}{\degree} elevation angle) for different algorithms: (a) original radar image, (b) proposed background removal algorithm, (c) mean subtraction algorithm, and (d) PCA algorithm.}
  \label{fig:rng_agl_maps_exp}
\end{figure*}

\subsubsection{Configuration}
The specific configuration of the cascaded-chip radar used in the data capture is as follows: center frequency \( f_{\R{c}} = \SI{77}{GHz} \), sweep rate \( S_{\R{w}} = \SI{45}{MHz \per\micro\second} \), sampling frequency \( f_{\R{s}} = \SI{15}{Msps} \), number of transmitters \( N_T = 12 \), number of receivers \( N_R = 16 \), chirp duration \( T_{\R{c}} = \SI{20}{\micro\second} \), sweep bandwidth \SI{384}{MHz}, the number of chirps per TX per frame is \( 128 \), the number of samples per chirp is \( 128 \), and the frame rate is \SI{30}{fps}.

\subsection{Experiments Scenario and Implementation}
\subsubsection{Scenario}
The experimental data was collected in a practical scenario with the ego vehicle in motion within a complex environment, as shown in Fig.~\ref{fig:imag_exp}. The scene consisted of various stationary objects such as fences, trees, buildings, and parked cars on the roadsides. Additionally, a bus was moving towards the ego vehicle. The testing phase involved analyzing 15 frames of recorded radar I-Q samples, capturing the dynamics of the scene and the interactions between the ego vehicle and its surroundings.

\subsubsection{Baseline}
We chose two baselines for comparison. The first baseline is the mean subtraction algorithm \cite{5505008}, which is used to remove strong reflections from the radar data by subtracting the mean signal computed by averaging multiple chirps. The second baseline is the principal component analysis (PCA) algorithm \cite{9337322}, which is employed for clutter removal by eliminating the first principal component of the signal that corresponds to background clutter. We did not consider any STAP methods as baselines, as it is challenging to obtain clutter-only training samples from data collected by a moving vehicular radar.

\subsubsection{Implementation}
We implemented the proposed 3D radar ego-motion estimation algorithm and static background removal algorithm using the following parameter settings. The Range, Doppler, Azimuth, and Elevation FFTs all employed 128 points. The CFAR false alarm probability was set to $10^{-2}$. The ambiguity set for the algorithm was ${\cal K}=\{-1, 0, 1\}$. For the 3D radar ego-motion estimation, we utilized the RANSAC algorithm with a sample size of 4, a maximum distance threshold of \SI{0.2}{m/s} for determining inliers, and a maximum number of trials set to 2000. Additionally, to remove strong reflections from the vehicle itself, we applied a mean subtraction in the Doppler domain.

\subsection{Background Removal Results}
We present the evaluation results of the proposed background removal algorithm and two baselines in Fig.~\ref{fig:dop_agl_maps_exp} and \ref{fig:rng_agl_maps_exp}. Our focus is on the plane of \SI{0}{\degree} elevation for the Doppler-azimuth angle image (Fig.~\ref{fig:dop_agl_maps_exp}) and the range-azimuth angle image (Fig.~\ref{fig:rng_agl_maps_exp}).

In Fig.~\ref{fig:dop_agl_maps_exp}(a), we observe that the Doppler components of the stationary background reflections span a range \([\SI{1.5}{m/s}, \SI{3}{m/s}]\), forming a U-shaped relationship with the azimuth angle. Considering the vehicle's average speed (which exceeds \SI{4}{m/s}), the actual Doppler velocity for the stationary background should be a negative value calculated based on the azimuth angle and radar ego velocity. The measured Doppler velocity within the range of \SI{1.5}{m/s} to \SI{3}{m/s} indicates the presence of Doppler ambiguity (i.e., \(k \neq 0\)). By utilizing our proposed 3D ego-motion estimation algorithm, the best \(k\) value is estimated as 1, confirming our earlier assumption. In Fig.~\ref{fig:dop_agl_maps_exp}(b), we observe that after applying the background removal with the proposed algorithm, the strong Doppler components for each azimuth angle are mostly filtered out, while the mainlobe from the moving bus and some sidelobe components from the background are still present. The mean subtraction baseline is ineffective at removing static clutter when the radar is moving, as it only removes the zero Doppler frequency, as shown in Fig.~\ref{fig:dop_agl_maps_exp}(c). The PCA algorithm removes most of the components, which also affects the moving objects, as shown in Fig.~\ref{fig:dop_agl_maps_exp}(d).

Furthermore, we assess the performance of the background removal algorithm on the range-azimuth angle image, as depicted in Fig.~\ref{fig:rng_agl_maps_exp}. In Fig.~\ref{fig:rng_agl_maps_exp}(a) and (b), we observe that the background components are efficiently removed with the proposed background removal algorithm, with the post-removal clutter amplitude below \SI{-10}{dB}, while the signal reflections from moving objects exhibit minimal change. In contrast, the mean subtraction shows little change to the whole image, and PCA's results are blurred significantly, as shown in Fig.~\ref{fig:rng_agl_maps_exp}(c) and (d). This demonstrates the effectiveness of the proposed algorithm in suppressing clutter and preserving the desired moving targets.

\begin{figure}
\centering
\includegraphics[width=0.42\textwidth]{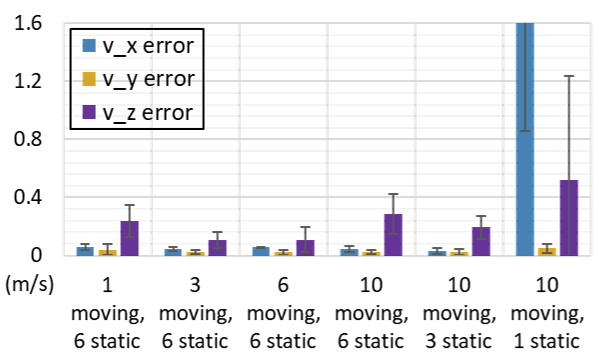}
\caption{Performance of the 3D ego-motion estimation algorithm under different scenarios with varying numbers of moving and static cars.}
\label{fig:sen_ego}
\end{figure}

\section{Discussion}
\subsection{Sensitivity Analysis for Ego-motion Estimation}
The performance of the proposed end-to-end background removal algorithm relies heavily on the accuracy of the 3D radar ego-motion estimation results. It is important to note that the RANSAC algorithm is not guaranteed to identify the correct inliers in all cases. Consequently, the 3D radar ego-motion estimation algorithm may converge to a model containing false inliers (i.e., data points that do not belong to the static background) or may miss some static elements, especially in challenging scenarios with few static objects and a high density of moving objects. In the former cases, the presence of false inliers in subsequent processing steps can lead to inaccurate background removal.

To illustrate the sensitivity of the ego-motion estimation algorithm, we extended the simulation to different scenarios. The original simulation scenario described in Section~\ref{sec_simu_scenario} included 6 static cars off the road and 1 moving car on the road (excluding the ego car). We increased the number of moving cars to 3, 6, and 10, respectively, and evaluated the performance of ego-motion estimation. The results, shown in Fig.~\ref{fig:sen_ego}, indicate that the ego-motion estimation errors for scenarios with 1, 3, 6, and 10 moving cars are on a similar level. This suggests that with a sufficient number of static cars, the ego-motion estimation remains robust to the number of moving objects. Next, we kept 10 moving cars and reduced the number of static cars to 3 and then to 1. The results, also shown in Fig.~\ref{fig:sen_ego}, demonstrate that ego-motion estimation performs well with 3 static cars but fails with only 1 static car. Therefore, when there are very few stationary objects relative to the number of moving targets, the RANSAC algorithm struggles to accurately identify the static background, affecting the performance of subsequent processing steps.

In summary, these findings indicate that the ego-motion estimation algorithm is generally robust, provided there are enough stationary objects in the scene. In scenarios with very few static objects relative to moving targets, incorporating additional methodologies such as sensor fusion might be necessary to enhance the robustness of 3D ego-motion estimation.

\begin{table}
    \centering
    \caption{The Time Complexity of Algorithms}
    \label{tab:time_cplx}
    \begin{tabular}{p{2.7cm}c}
       \toprule
         Methods &  Time Complexity\\
        \midrule
        Proposed algorithm  & $\mathcal{O}(N_sN_cN_hN_e\log{N_sN_cN_hN_e})$ \\
        \midrule
        Mean subtraction \cite{5505008} & $\mathcal{O}(N_sN_cN_hN_e)$\\
        \midrule
         PCA \cite{9337322} & $\mathcal{O}((N_sN_c^2 + N_c^3)N_hN_e)$ \\
       \bottomrule
    \end{tabular}
\end{table}

\subsection{Complexity Analysis}
We now analyze the time complexity of the proposed background removal algorithm and the two baseline algorithms used for comparison. Let the dimensions of the radar data cube be denoted as $N_s$ (samples), $N_c$ (chirps), $N_h$ (horizontal antennas), and $N_e$ (vertical antennas). We assume that FFT operations on any dimension do not change the input data size.

The overall time complexity of the proposed algorithm can be divided into three parts: 3D ego-motion estimation, radar imaging, and background filtering. The approximate complexity of 3D ego-motion estimation is $\mathcal{O}(N_sN_cN_hN_e\log{N_sN_c}) + \mathcal{O}(N_sN_c)$, dominated by the range-Doppler FFT and CFAR detection operations. The complexities of the azimuth and elevation angle FFTs, as well as the LSR estimator, are not included, as the number of CFAR detections in these steps is relatively small. The complexity of radar imaging is approximately $\mathcal{O}(N_sN_cN_hN_e\log{N_hN_e})$, assuming that the range-Doppler intermediate results from the previous step are used. The complexity of background filtering is $\mathcal{O}(N_hN_eC)$, where $C$ is the computation required for each notch filter. Based on prior analysis, $C$ is very small. Therefore, the total time complexity of the proposed algorithm is $\mathcal{O}(N_sN_cN_hN_e\log{N_sN_cN_hN_e})$, ignoring smaller terms.

For the mean subtraction algorithm \cite{5505008}, the time complexity is $\mathcal{O}(N_sN_cN_hN_e)$, driven by the mean subtraction across chirps. The PCA algorithm \cite{9337322} has a time complexity of $\mathcal{O}((N_sN_c^2 + N_c^3)N_hN_e)$, as derived from \cite{yigit2024time}.

\section{Conclusion}
This paper introduces an efficient algorithm for static background removal in automotive radars, utilizing 4D radar imaging and filtering within the azimuth-elevation-Doppler domain. Extensive evaluations demonstrate the algorithm's effectiveness in suppressing background clutter and optimizing computational efficiency. Future work will focus on enhancing the accuracy of 3D ego-motion estimation, as the performance of background removal depends critically on this accuracy. By improving ego-motion estimation, we aim to further enhance the algorithm’s performance and robustness across various real-world applications.


\bibliographystyle{IEEEtran}
\bibliography{bibtex}

\end{document}